\documentclass[12pt,a4paper,twoside]{paper}
\usepackage{amsmath,bbm,amssymb,amsxtra}
\usepackage{hyperref}
\usepackage{graphicx}
\usepackage{epstopdf}
\usepackage{chngcntr}
\usepackage{etoolbox} 
\usepackage{times}
\def\R{\mathbb R}

\newcommand{\NX}{{{\bf X}}}

\newcommand{\Nx}{{{\bf x}}}
\newcommand{\Ny}{{{\bf y}}}

\ifx\optionkeymacros\undefined\else \fi

\catcode`\Œ=\active\defŒ{{\aa}}       
\catcode`\º=\active\defº{\int}        
\catcode`\=\active\def{\c c}        
\catcode`\¶=\active\def¶{\partial}    
\catcode`\Ä=\active\defÄ{\oint}       
\catcode`\Æ=\active\defÆ{\triangle}   
\catcode`\Â=\active\defÂ{\neg}        
\catcode`\µ=\active\defµ{\mu}         
\catcode`\¿=\active\def¿{{\o}}        
\catcode`\¹=\active\def¹{\pi}         
\catcode`\Ï=\active\defÏ{{\oe}}       
\catcode`\§=\active\def§{{\ss}}       
\catcode`\ =\active\def {\dagger}     
\catcode`\Ã=\active\defÃ{\sqrt}       
\catcode`\·=\active\def·{\Sigma}      
\catcode`\Å=\active\defÅ{\approx}     
\catcode`\½=\active\def½{\Omega}      
\catcode`\£=\active\def£{{\it\$}}     
\catcode`\°=\active\def°{\infty}      
\catcode`\¤=\active\def¤{{\S}}        
\catcode`\¦=\active\def¦{{\P}}        
\catcode`\¥=\active\def¥{\bullet}     
\catcode`\»=\active\def»{\leavevmode\raise.585ex\hbox{\b a}}      
\catcode`\¼=\active\def¼{\leavevmode\raise.6ex\hbox{\b o}}        
\catcode`\­=\active\def­{\not=}       
\catcode`\²=\active\def²{\leq}        
\catcode`\³=\active\def³{\geq}        
\catcode`\Ö=\active\defÖ{\div}        
\catcode`\É=\active\defÉ{{\dots}}     
\catcode`\¾=\active\def¾{{\ae}}       
\catcode`\Ç=\active\defÇ{\ll}         
\catcode`\Ò=\active\defÒ{``}          
\catcode`\Á=\active\defÁ{!`}          
\catcode`\¢=\active\def¢{\rlap/c}     
\catcode`\Ô=\active\defÔ{`}           
\catcode`\Õ=\active\defÕ{'}           

\catcode`\=\active\def{{\AA}}       
\catcode`\'=\active\def'{\c C}        
\catcode`\¯=\active\def¯{{\O}}        
\catcode`\¸=\active\def¸{\Pi}         
\catcode`\Î=\active\defÎ{{\OE}}       
\catcode`\®=\active\def®{{\AE}}       
\catcode`\×=\active\def×{\diamond}    
\catcode`\¡=\active\def¡{\accent'27}  
\catcode`\Ó=\active\defÓ{''}          
\catcode`\±=\active\def±{\pm}         
\catcode`\È=\active\defÈ{\gg}         
\catcode`\À=\active\defÀ{?`}          
\catcode`\Ð=\active\defÐ{--}          
\catcode`\Ñ=\active\defÑ{---}         

\catcode`\Š=\active\defŠ{\"a}        
\catcode`\'=\active\def'{\"e}        
\catcode`\•=\active\def•{\"{\i}}     
\catcode`\š=\active\defš{\"o}        
\catcode`\Ÿ=\active\defŸ{\"u}        
\catcode`\Ø=\active\defØ{\"y}        
\catcode`\€=\active\def€{\"A}        
\catcode`\…=\active\def…{\"O}        
\catcode`\†=\active\def†{\"U}        
\catcode`\‡=\active\def‡{\'a}        
\catcode`\Ž=\active\defŽ{\'e}        
\catcode`\'=\active\def'{\'{\i}}     
\catcode`\—=\active\def—{\'o}        
\catcode`\œ=\active\defœ{\'u}        
\catcode`\ƒ=\active\defƒ{\'E}        
\catcode`\ˆ=\active\defˆ{\`a}        
\catcode`\=\active\def{\`e}        
\catcode`\"=\active\def"{\`{\i}}     
\catcode`\˜=\active\def˜{\`o}        
\catcode`\=\active\def{\`u}        
\catcode`\Ë=\active\defË{\`A}        
\catcode`\‹=\active\def‹{\~a}        
\catcode`\–=\active\def–{\~n}        
\catcode`\›=\active\def›{\~o}        
\catcode`\Ì=\active\defÌ{\~A}        
\catcode`\"=\active\def"{\~N}        
\catcode`\Í=\active\defÍ{\~O}        
\catcode`\‰=\active\def‰{\^a}        
\catcode`\=\active\def{\^e}        
\catcode`\"=\active\def"{\^{\i}}     
\catcode`\™=\active\def™{\^o}        
\catcode`\ž=\active\defž{\^u}        

\let\optionkeymacros\null

\begin{document}
   
     \begin{center}
{\Huge Why Bohm and Only Bohm?\footnote{Dedicated to the memory of Detlef D\"urr}}  \vspace*{5mm}

 \vspace*{5mm}
{\LARGE Jean Bricmont
 
 IRMP, UCLouvain, chemin du Cyclotron 2, 1348 Louvain-la-Neuve
Belgium}\footnote{E-mail: jean.bricmont@uclouvain.be}
 \vspace*{5mm}

 \hspace*{70mm} {\Large\it  Particles move}\footnote{Detlef's answer to a famous mathematical physicist with no interest in foundations of QM who
was asking him what Bohm's theory was all about ``in two words".} 

\end{center}
\begin{abstract}

 It is often claimed that there are three ``realist" versions of quantum mechanics: the de Broglie-Bohm theory or Bohmian mechanics, the spontaneous collapse theories and the many worlds interpretation.

We will explain why the two latter proposals suffer from serious defects coming from their ontology (or lack thereof) and that the many worlds interpretation is unable to account for the statistics encoded in the Born rule.  The de Broglie-Bohm theory, on the other hand, has no problem of ontology and accounts naturally for the Born rule.
\end{abstract}
 
 \newpage
\section{A Misleading Problem: The Measurement One}\label{sec1}

The measurement problem is well known: at the end of an experiment where one measures the property of a particle that can take two values, the wave function (or quantum state) of the measuring device, or of the cat if we couple the device to the cat through a poison capsule, is (leaving aside normalization factors):
 
\begin{equation}
\Psi_{\mbox{cat alive}}+\Psi_{\mbox{cat dead}}.
\nonumber	
\end{equation}

 And that cannot be a complete description of the cat, which is obviously either alive or dead but not both!
  
  The way out of this problem from the point of view of ordinary quantum mechanics is to introduce the collapse postulate: when one looks at the cat, one sees whether she is alive or dead and, depending on what one sees, one reduces the wave function of the cat (and of the particle that was measured and is thus coupled to the state of the cat) to either $\Psi_{\mbox{cat alive}}$ {\it or} $\Psi_{\mbox{cat dead}}$.

  Since this is a {\it deus ex machina} from the point of view of the linear Schr\"odinger evolution, justifying it is often viewed as the main problem in foundations of quantum mechanics.

  However there is a deeper problem: neither $\Psi_{\mbox{cat alive}}$ nor $\Psi_{\mbox{cat dead}}$ are cats: they are functions defined on a high dimensional space $\R^{3N}$  while cats are located in $\R^3$. And it is not clear what it means to say that $\Psi_{\mbox{cat alive}}$ or $\Psi_{\mbox{cat dead}}$ are descriptions of cats, let alone ``complete descriptions" of them\footnote{This idea is emphasized by R. Tumulka in \cite[Sect. 5.1]{Tu2}.}.  	
 
What most people do is to mentally identify cats and wave functions of cats, which is illegitimate. So, the real problem is the meaning or the ontology one: what does the wave function mean outside of laboratories, and what does it say about what the world is made of?

 \section{ An Intuitive Solution That Does Not Work}\label{sec2}

 A simple and a priori attractive solution to this problem is the
na\"ive statistical interpretation of quantum mechanics (and that is probably what is in the back of the minds of most of the ``no worry about quantum mechanics" physicists): particles do have properties such as position, velocity, spin etc., but we cannot know or control them-we have only access to their wave function.

 That object gives the statistical distribution of the values of those quantities (through the Born rule) over sets of particles having the same wave function. And, when we perform a measurement of a property of a given  particle, we learn what that value is for that particle.

 In that interpretation, the reduction of the wave function is no problem; we simply adjust our probabilities when we learn something new about the system.

And, if that worked, there would indeed be no reason to worry about the meaning of the wave function and we would have a decent meaning of that function outside of measurements. However, it cannot work, because of well-established theorems due to John Bell \cite{Be1} and to Simon Kochen and Ernst Specker \cite{KS}, but that unfortunately are not widely known among physicists.

 Those theorems show that, if we assume that there exists a map $v$ that assigns a value to each observable $A$ corresponding to various properties that are simultaneously measurable according to ordinary quantum mechanics and that agrees with minimal quantum mechanical predictions concerning such observables, then one can deduce a contradiction\footnote{See Mermin \cite{Me4} for pedagogical proofs and \cite{BGH,DDGZ} for more details.}.

 These theorems are called the ``no hidden variables theorems".

 Obviously, there cannot be a statistical distribution of maps that do not exist. In other words, what we call the statistical interpretation (individual quantum systems do have definite properties but we are only able to know their statistical distributions) does not work.

Note in passing that the ideas behind these no hidden variables theorems can be used to rule out the ``decoherent histories" approach of M. Gell-Mann, J. Hartle, R. Griffiths and R. Omn\`es \cite{GMH,Gri,Om}.

 Indeed, as noticed by S. Goldstein \cite{Go} and A. Bassi and G. Ghirardi \cite{BG1}, this approach amounts to assigning simultaneous values to pairs of commuting observables; but if a series of such pairs is suitably chosen, a contradiction follows.

 So, let us look for non obvious solutions.

\section {Spontaneous Collapse Theories}\label{sec3}

These theories are modified versions of quantum mechanics, the first of which was  
 introduced by  Ghirardi, Rimini, and Weber \cite{GRW}  in which wave functions spontaneously collapse\footnote{For reviews and further discussions of those theories, see  Ghirardi \cite{G},  Ghirardi  {\it et al} \cite{BGG}, Allori  {\it et al} \cite{ASTZ}, Goldstein {\it et al} \cite{GTZ}.}.

 To be precise, in that model, the wave function evolves according to the Schr\"odinger equation most of the time, but there is a set of spacetime points $(\Ny_i, t_i)$ chosen at random, such that the wave function $\Psi (\Nx_1, \dots, \Nx_N, t) $ for a system of $N$ particles is multiplied at the chosen times $t_i$ by a Gaussian function in the variable $\Nx_k$ ($k$ chosen uniformly among $1, \dots, N$), centered in space at the chosen space points $\Ny_i$. 
	
The probability distribution of these random points is determined by the wave function of the system under consideration at the times when they occur, and is given by the familiar $|\Psi|^2$ distribution. This ensures that the predictions of the GRW theory will (almost) coincide with the usual ones.

The above-mentioned multiplication factors localize the wave function in space, and, for a system of many particles in a superposed state, effectively collapse the wave function onto one of the terms. Now, the trick is to choose the parameters of the theory so that spontaneous collapses are rare enough for a single or for a few particles in order to ensure that they do not lead to detectable deviations from the quantum predictions, but  are frequent enough to ensure that a system composed of  a large number of particles, say $N=10^{23}$,  will not stay in a superposed wave function  for more than a split second.

Spontaneous collapse theories are not the same as ordinary quantum mechanics, since they lead to predictions that differ from the usual ones, even for systems made of a small number of particles. But the parameters of the theory are simply adjusted so as to avoid being refuted by present experiments, which is not exactly an appealing move.

Moreover there is the problem of making sense of a pure wave function ontology (even when the latter collapses, since the collapsed wave function is still just a function defined on a high-dimensional space).

Two solutions have been proposed to give a meaning to the GRW theory beyond the pure wave function ontology: the matter density ontology  often denoted GRWm \cite{BGG, ASTZ}, and the flash ontology  denoted GRWf \cite{Be8}.

The matter density ontology associates a continuous matter density to the wave function of a system of $N$ particles. For each $\Nx\in \R^3$, and $t\in \R$, one defines:
\begin{equation}
m(\Nx, t)= 
\sum_{i=1}^N m_i \int_{\R^{3N}} \delta(\Nx-\Nx_i) |\Psi (\Nx_1, \dots, \Nx_N, t)|^2 d \Nx_1 \dots d\Nx_N\;, \label{1}
\end{equation}
 where $\Psi (\Nx_1, \dots, \Nx_N, t)$ is the usual wave function of the system at time $t$. This equation makes a connection between the wave function defined on the high-dimensional configuration space and an object, the matter density, existing in our familiar space $ \R^3$.

 In our three-dimensional world, there is just a continuous density of mass:  no structure, no atoms, no molecules, etc., just an amount of ``stuff", with high density in some places and low density elsewhere.

In the flash ontology, one has a world made only of spacetime points at the center of the Gaussian  multipliers of the wave function that collapse it. No particles, no fields, nothing at all, except a ``galaxy" of spacetime points, called ``flashes".
	
	Let us note that, if the God of the physicists was trying not to be malicious and if either the matter density or the flash ontologies are   true, then He failed badly: indeed, it means that we were wrong all along when we ``discovered" atoms, nuclei, electrons, etc., and that we are lying to schoolchildren when we tell them that matter is mostly void with a few pieces of matter (the atoms) here and there. Indeed, in the matter density ontology, matter is continuous after all, with higher and lower density in some places, and we have simply been fooled by this modified version  of quantum mechanics into thinking that it is not.

On the other hand, if the flash ontology is true, then we have been fooled into thinking that there exists something most of the time (like atoms): if we take the visible universe since the Big Bang,  it has contained only finitely many flashes. Since the flashes are all there is in that ontology, this means that, most of the time, the universe is just empty.

Of course, even if those ontologies are weird,  we might be forced to accept one of them (after all the existence of atoms in  empty space is also counterintuitive),
 if there were independent reasons for doing so, like a greater explanatory power or greater empirical adequacy.

 But, and this is the most important point, empirical adequacy of any spontaneous collapse theory would mean that ordinary quantum mechanics is empirically wrong, since the predictions of both theories differ, at least in principle. So, if one found that a prediction of a spontaneous collapse theory is right, when it differs from ordinary quantum mechanics, it would be a major revolution in physics and we might be forced to worry about those weird ontologies. But this hasn't happened yet and I would suggest not to worry about what to do after the revolution before that revolution has occurred.

To illustrate how odd the continuous matter and the flash ontologies  are, compared to the particle ontology of the de Broglie--Bohm theory (to be discussed in Sect.~\ref{sec5}), consider what happens, in those ontologies, with the thought experiment of Einstein's boxes: imagine a box containing just one particle that is cut in two parts; one part is  sent to New York, the other to Tokyo \cite{Boxes}, \cite[Chap. 10]{Ma}. 

\begin{figure}[t]
\centering
\includegraphics[width=.8\textwidth]{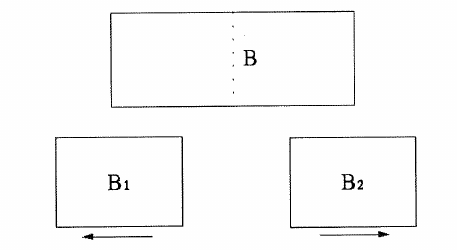}
\caption[]{Einstein's boxes. Reproduced with permission from T. Norsen \cite{Boxes}. Copyright  2005 American Association of Physics Teachers}\label{4fig1}
\end{figure}

The wave function of the particle is a superposition of a wave function located in box $B_1$ $+$ a wave function located in  box $B_2$. In ordinary quantum mechanics, when Alice in New York opens her half-box and sees the particle, the wave function collapses on the part of the wave function located in New York and, if she doesn't see it, the wave function collapses on the part located in Tokyo.

In the de Broglie--Bohm theory, nothing surprising happens: the particle is in one of the half-boxes all along and is found where it is.

But in the GRWm theory, there is one-half of the matter density of a single particle in each half-box. When Alice couples  her half-box with a detector of particles, the evolution of the wave function of the particle is coupled with a macroscopic object and many collapses occur quite rapidly, so that the matter density suddenly jumps from being one-half of the matter density of a single particle in each half-box to being the full matter density of a particle in  that half-box and nothing in the other.

There is a nonlocal transfer of matter in the GRWm theory, while there is no such thing in the de Broglie--Bohm theory, and not even anything nonlocal when one deals with only one particle.

 In the GRWf theory, there is simply nothing in either half-box, just a wave function traveling so to speak with the half-boxes. When Alice couples  her half-box with a detector of particles, the wave function of the particle becomes coupled with a macroscopic object, and there is suddenly a ``galaxy of flashes" appearing (randomly) in that detector, either detecting the particle or not; we interpret the first possibility as meaning that the particle is in that half-box and the second one as meaning that the particle is in the other half-box.

Putting aside the weirdness of the ontologies, the spontaneous collapse theories are more nonlocal than the EPR-Bell result implies, since the latter concerns nonlocality for systems with at least two particles.

 Here we have nonlocal effects or actions at a distance (Alice affects the physical situation in Tokyo by acting in New York) {\bf even for one particle}!

 In summary:
 
\begin{itemize} 

\item Spontaneous collapse theories have unnatural ontologies.

\item They are very ad hoc: parameters are chosen so as to avoid refutation and not on the basis of any evidence\footnote{Moreover, recent results seem to provide evidence refuting spontaneous collapse theories: https://www.quantamagazine.org/physics-experiments-spell-doom-for-quantum-collapse-theory-20221020/}.

\item They can only be true if quantum mechanics itself is false.

\item They are more nonlocal than they have to be.
\end{itemize}

\section{The Many-Worlds Interpretation}\label{sec4}

This interpretation postulates that, when the proverbial cat (or any other macroscopic device) finds itself in a superposed state, then, instead of undergoing a collapse by fiat as in ordinary quantum mechanics, both terms simply continue to exist. But how can that be possible? We always see the cat alive {\it or\/} dead but not both! The short answer is that they both exist, but in different ``worlds".

Hence, whenever an experiment leads to a macroscopic superposition, the world splits into two or more worlds, depending on the number of distinct macroscopic states produced by that experiment, one for each possible result.

Why do I always perceive only one of the results? It is simple: I, meaning my body, my brain (and thus also my consciousness) becomes entangled with the states of the cat, so there are two or more copies of me also, one seeing the dead cat in one world, another seeing the live cat in another world. And that, of course, is also true for everything else: every molecule in the entire world becomes copied twice (maybe not instantaneously, but that is another question).

In his original paper \cite{Eve}, Everett stressed that ``{\it all\/} elements of a superposition are `actual', none any more `real' than the rest." Everett felt obliged to write this because ``some correspondents'' had written to him saying that, since we experience only one element of a superposition, we have only to assume the existence of that unique element. This shows that some early readers of Everett were already baffled by the radical nature of the ``many-worlds" proposal.

Putting aside  the weirdness of this multiplication of ``worlds", one should ask whether the  many worlds scheme is coherent. Consider  the Born rule. Suppose that the probabilities  of having the cat alive or dead, as a result of an experiment, are $(\frac{1}{2}, \frac{1}{2})$. And suppose that I decide to repeat the same experiment successively many times, with different particles (and cats) but all having the same initial wave function.

After one experiment, there are two worlds, one with a dead cat and a copy of me seeing a dead cat and one with a live cat and a copy of me seeing a live cat. Since both copies of me are in the same state of mind as I was before the first experiment (after all, both copies are just copies of me!), each of them repeats that experiment.

Then, we have four worlds, one with two consecutive dead cats, one with two consecutive live cats and two with one  dead cat and one live cat. ``I" (by that I mean each copy of me in each of those four worlds) repeat the experiment again: we have now eight worlds, with one ``history" of worlds with three dead cats, 
one history of worlds with three live cats, three histories of worlds with one live cats and two dead ones, three histories of worlds with one dead cat and two live ones.

Now, continue repeating that experiment: 
 for every possible sequence of outcomes, there will be some of my ``descendants" (i.e.  copies of me, that exist in all the future worlds) that will see it. There will be a sequence of worlds in which the cats are always alive and another sequence where they are always dead. There are also many sequences of worlds where the cats are alive one quarter of the time and dead three quarters of the time, and that is true for any other statistics different from $(\frac{1}{2}, \frac{1}{2})$. So that we can be certain that many of our descendants will {\it not} observe Born's rule in their worlds.

But one could argue, on the basis of the law of large numbers, that, at least in the vast majority of worlds, the Born rule will be obeyed, since, in the vast majority of worlds, the frequencies of dead and live cats will be close to $(\frac{1}{2}, \frac{1}{2})$.

But what happens if, instead of being $(\frac{1}{2}, \frac{1}{2})$, the probabilities predicted by quantum mechanics are, say, $(\frac{3}{4}, \frac{1}{4})$? We will still have two worlds coming out of each experiment, because these experiments have two possible outcomes. So, the structure of the multiplication of worlds is exactly the same as when the predicted probabilities were $(\frac{1}{2}, \frac{1}{2})$.

But now, if one applies the law of large numbers as above,
one arrives at the conclusion that, in the vast majority of worlds, the quantum predictions will {\it not\/} be observed, since our descendants will still see the cats alive in approximately  $\frac{1}{2}$ of the worlds, and the cats dead also  in approximately  $\frac{1}{2}$ of the worlds,
 instead of the $(\frac{3}{4}, \frac{1}{4})$ frequencies predicted by the Born rule.

This is  a serious problem for the many-worlds interpretation. There have been many proposals to solve this problem and it would be too long and too technical to discuss all of them here.

Some authors have argued that one should count the worlds differently, by weighting them with the coefficients that appear in the Born rule \cite{Eve,DWG}.

 However, this does not answer the objection above.

Another "solution" is to give to low probability worlds (according to Born's rule) a lower degree of existence or of reality, but it is unclear what it means to live in a low reality world since we cannot compare that life with one in a world with a high degree of existence: indeed, different worlds don't interact with each other.

And this ``solution" is quite contrary to Everett's original idea that ``{\it all\/} elements of a superposition  are `actual', none any more `real' than the rest".

Moreover, one also has the problem of ontology: what proliferates are wave functions, but the latter are mathematical objects not ``worlds" in space-time.

One solution is to associate to the wave function 
 the continuous matter density (\ref{1}), as in the spontaneous collapse theories, see \cite{ASTZ1}.

However, this does not solve the problem of the probabilities discussed above. Coming back to our example with two possible outcomes, one having probability $\frac{3}{4}$ and the other $\frac{1}{4}$, the density of matter will be different in the world where one sees the outcome having probability $\frac{3}{4}$ from the one where one sees the outcome having probability $\frac{1}{4}$.

But what difference does it make? In which way does having a smaller or larger matter density affect my states of mind ? And if it does not, we are back to the problem that, if one repeats many times the experiment whose outcomes have probabilities $(\frac{3}{4}, \frac{1}{4})$, most of my descendants (some of course having a small matter density) will see massive violations of the Born rule.
 
\section{The de Broglie-Bohm Theory}\label{sec5}

\hspace*{20mm} {\it Nature and Nature's Laws lay hid in Copenhagen: God said, "Let de Broglie-Bohm be!" and all was light.}\footnote{Adapted from Alexander Pope's epitaph about Newton.}

In the de Broglie-Bohm theory, the complete state of a system with $N$ variables at time $t$ is specified by $\big(\Psi (t), {\bf X}(t)\big)$, where $\Psi (t)$ is the usual wave function, $\Psi (t)=\Psi(\Nx_1, \dots, \Nx_N, t)$ and ${{\bf X}(t)}= \big(\NX_1(t),\ldots, \NX_N (t)\big)\in\R^{3N}$ are the actual positions of the particles\footnote{Our presentation of the de Broglie-Bohm theory follows the one of Bell \cite{B} and of  D\"urr, Goldstein and Zangh\`i \cite{DGZ1} rather than the one of Bohm \cite{Bo1}. Many expositions of the de Broglie--Bohm theory are available, see, e.g., Albert \cite{Al1} or Tumulka \cite{Tu1} for elementary introductions and, e.g.  Bohm and Hiley \cite{BH}, Bricmont (\cite{Bri1}, D\"urr and Teufel \cite{DT},  Goldstein \cite{Go-St}, Maudlin \cite{Ma2}, Norsen \cite{No} and Towler \cite{Tow} for more advanced ones.}.

The theory assumes that the particles have positions at all times, and therefore trajectories, independently of whether one measures them or not.

The evolution  of the state ($ \Psi , \bf X$) is given by two laws:
\begin{enumerate}
\item $\Psi$ obeys the usual Schr\"odinger equation at all times. The wave function of an isolated system never collapses.

\item The evolution  of the positions of the particles is guided by the wave function at time $t$. 
 
\end{enumerate}

Moreover, since the system is deterministic, one has to make statistical assumptions on the initial conditions  of the system (as one does in chaos theory) in order to obtain statistical predictions.

These are the fairly natural ``quantum equilibrium" ones: $\rho=|\psi|^2$.

Let me list the main qualities of this theory:

\begin{itemize}
	\item 

 Under the quantum equilibrium assumptions on initial conditions,  one recovers the usual predictions of quantum mechanics.
\item 
The ontology of the theory is the same as in classical physics and is thus unproblematic, unlike the ontologies of the spontaneous collapse and many worlds theories\footnote{One might replace particle's positions by field configurations in quantum field theories, but that goes beyond the scope of this article.}. It can be summarized by Detlef's quote mentioned at the beginning of this article: ``matter moves". And while in classical physics, that motion is guided by gravitational or electromagnetic fields, here it is guided by the wave function, a more abstract notion but a perfectly well-defined one.

\item 

The de Broglie-Bohm theory gives a clear physical meaning to the wave function, which is no longer simply a ``probability wave" (whatever that means exactly) but a physical quantity determining the motion of particles, similar in some ways to classical Hamiltonians.

\item 
 This theory explains the ``contextuality" of measurements.

Consider an idealized spin measurement: if the wave function has a symmetry along the $z$ axis and if the particle starts above the line of symmetry $z=0$, it will be deflected upwards, meaning that its spin is ``up", see figure \ref{fig5.1}.

\begin{figure}[!ht]
\centering
\includegraphics[width=.8\textwidth]{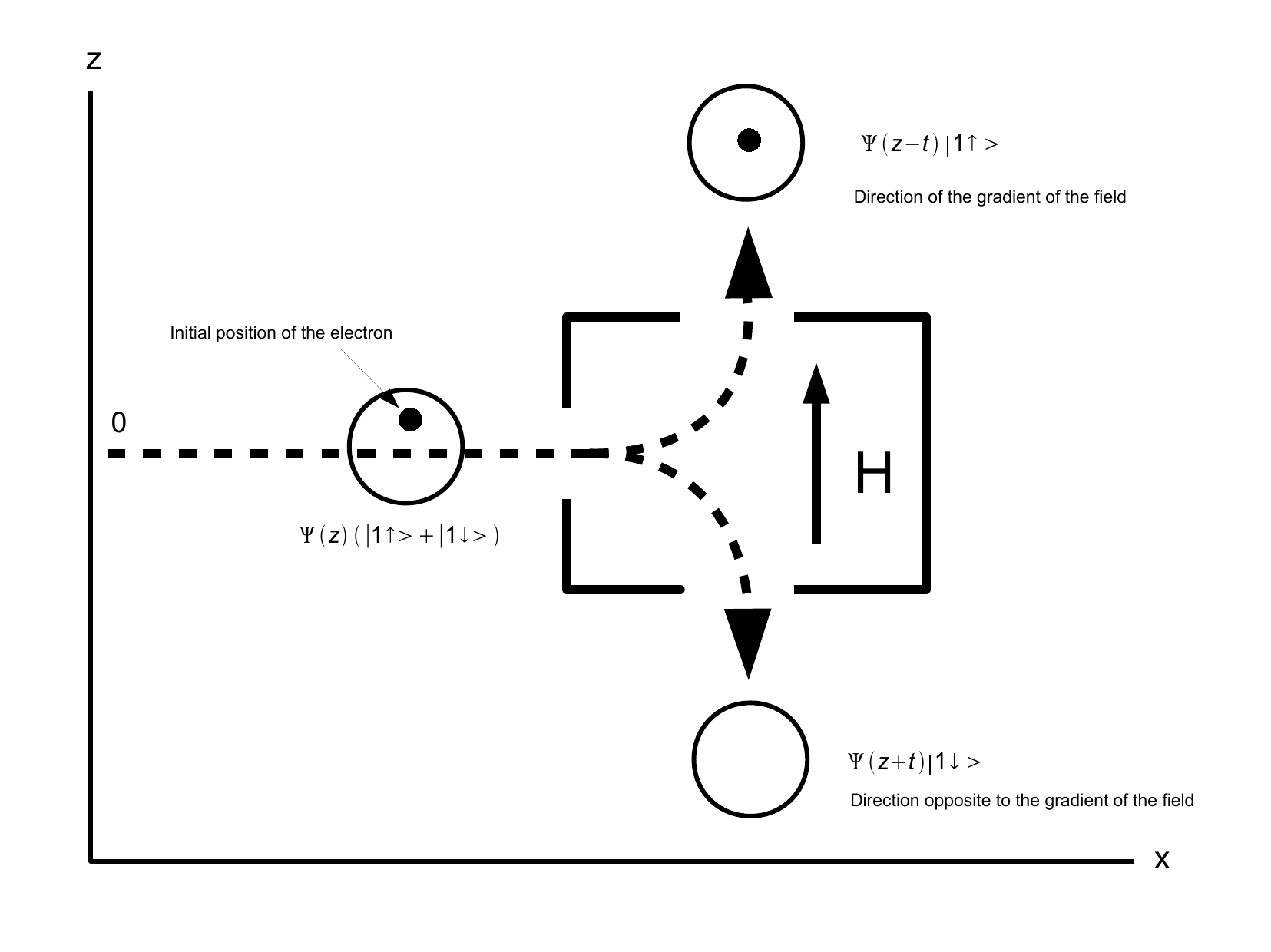}
\caption[]{An idealized spin measurement}\label{fig5.1}
\end{figure}

But if we reverse the orientation of the gradient of the magnetic field and do the same ``measurement" with the same initial wave function and the same initial position of the particle, the particle will still be deflected upwards, but that means now that its spin is ``down", see figure \ref{fig5.2}.

\begin{figure}[!ht]
\centering
\includegraphics[width=.8\textwidth]{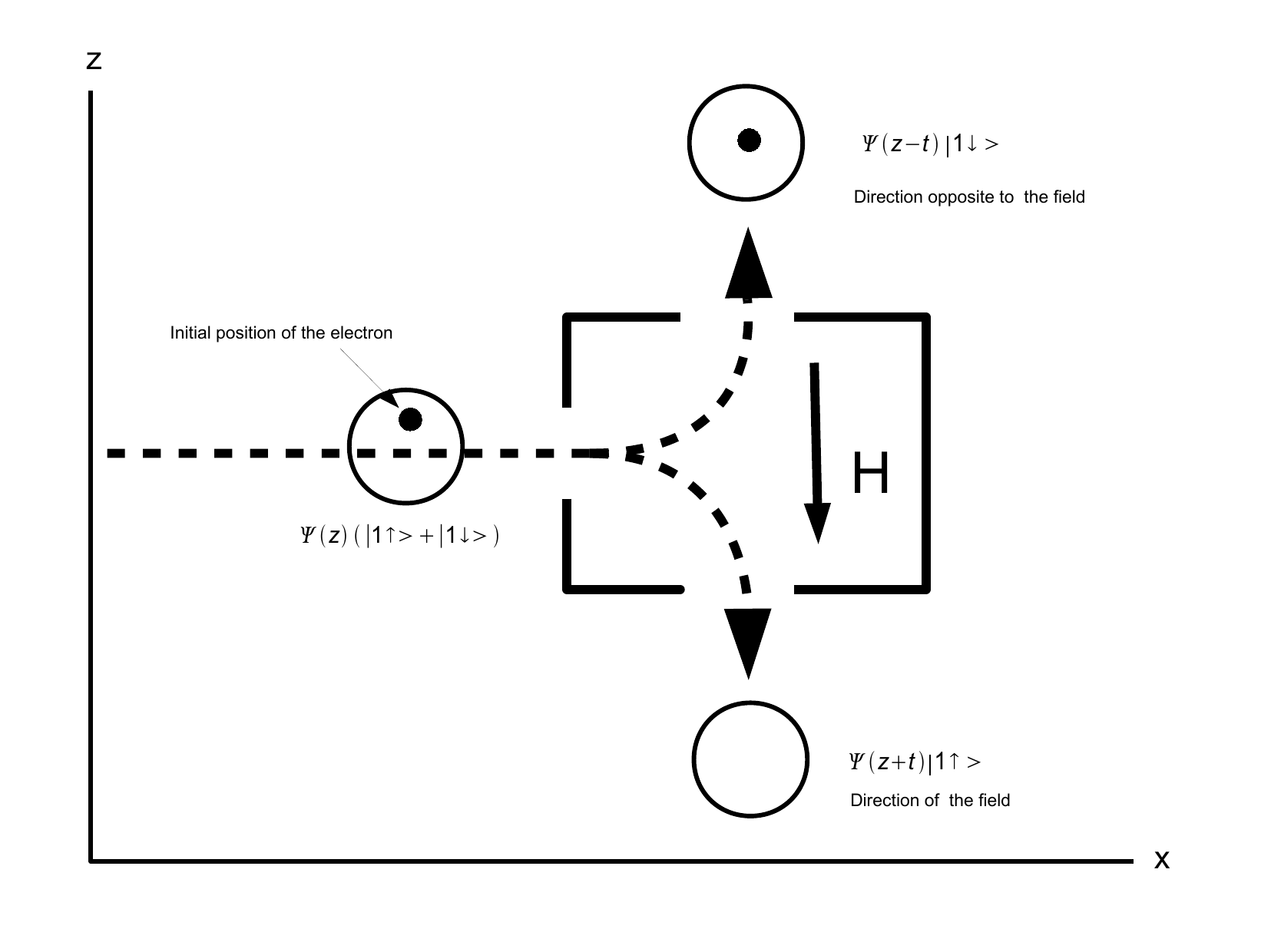}
\caption[]{An idealized spin measurement with the orientation of the gradient of the magnetic field reversed with respect to the one in figure \ref{fig5.1} }\label{fig5.2}
\end{figure}

So, ``measurements"	 don't measure  intrinsic properties of the system (except for position measurements).

Measurements are {\bf interactions} between an apparatus and a system.

This was stated intuitively by Bohr, but here it follows from the equations of the theory.

\item 
The de Broglie-Bohm theory is  a statistical theory, but, unlike the na\"ive one mentioned in the Sect.~\ref{sec2}, it is consistent and is not refuted by the no hidden variables theorems of Bell  and Kochen and Specker, because it does not introduce the ``hidden variables" that are forbidden by these theorems, like spin values preexisting to their ``measurement".

A subtle but crucial point: the de Broglie-Bohm theory is a hidden variable theory that is not refuted by the no hidden variables theorems.

Often missed (to put it mildly)!

\item 
One can use this contextuality of measurements to illustrate how nonlocality works in the de Broglie-Bohm theory.

 If the result of a ``measurement" on one side of an EPR-Bell experiment with an entangled pair of particles depends on how the orientation of the gradient of the field  is oriented on that side, then changing that orientation will affect the behavior of the particle on the other side: since the spins of both particles have to be anti-correlated, if changing that orientation  on one side changes the ``value" of the spin on that side from up to down, then the spin on the other side must go from down to up. But that means that the trajectory followed  by the particle on the other side must also change.

 This makes nonlocality explicit in the de Broglie-Bohm theory, which is a (big) quality rather than a defect since Bell has shown that nonlocality is a property of the world and not only of the quantum theory. 
\end{itemize}

\section{Conclusions}\label{sec7}

While we have left out certain questions such as quantum field theory and relativity (that can be dealt with within de Broglie-Bohm theory but it would be too long to discuss that), we want to stress certain aspects of that theory:

The de Broglie-Bohm theory is {\bf not} a different theory from ordinary quantum mechanics. It is the rational completion of ordinary quantum mechanics. The latter is just the algorithm allowing us to predict ``results of measurements" and that algorithm can be derived from the de Broglie-Bohm theory.

Most physicists either don't care  about the meaning of their most fundamental theory (``shut up and calculate") or adhere (in the back of their mind) to the na\"ive statistical interpretation.

Most physicists won't be persuaded by spontaneous collapse theories (too ad hoc) unless some future experiments contradict ordinary quantum predictions.

If that happens, they are likely to look for an entirely different theory (nonlinear?)

An important minority of physicists ``like" the Many-Worlds interpretation.

But I believe that this is because they haven't thought it through.

Apart from its fantastic nature, one has to provide it with an ontology and solve the problem of the statistical predictions (unsolved since 1957).

So, that leaves us with the de Broglie-Bohm theory as the only option. I believe it is gaining popularity, due in part to the work of Detlef D\"urr, although it is still very marginal.

 For it to become more popular, one needs:

\begin{itemize}
	\item 
 That physicists start to worry about the meaning of their most fundamental theory.
 
	\item That they be better aware of the nature of the problem: not the measurement one, but the meaning or ontology. And that they also become aware of the no hidden variable theorems that refute their  na\"ive statistical interpretation.

\end{itemize}

But that is a long way to go!

{\bf Acknowledgments} I thank  Sheldon Goldstein and Tim Maudlin for many discussions on Bohmian mechanics and the many-worlds theories.

\end{document}